\documentstyle[psfig]{mn}

\newcommand{\HI}{H\,{\sc i}}
\newcommand{\HII}{H\,{\sc ii}}
\newcommand{\Ha}{H$\alpha$}
\newcommand{\kms}{~km\,s$^{-1}$}
\newcommand{\kkms}{km\,s$^{-1}$}

\newcommand{\rHI}{$r_{\rm HI}$}
\newcommand{\vsys}{$v_{\rm sys}$}

\newcommand{\vrot}{$v_{\rm rot}$}
\newcommand{\FHI}{$F_{\rm HI}$}

\newcommand{\LB}{$L_{\rm B}$}
\newcommand{\LX}{$L_{\rm X}$}
\newcommand{\MHI}{$M_{\rm HI}$}
\newcommand{\Mtot}{$M_{\rm tot}$}

\newcommand{\Msun}{~M$_{\odot}$}

\newcommand{\Lsun}{~L$_{\odot}$}
\newcommand{\Ho}{H$_{\rm o}$}

\title[Neutral Hydrogen Gas in Interacting Galaxies]
      {Neutral Hydrogen Gas in Interacting Galaxies: \\
       The NGC~1511 galaxy group\thanks{The observations were obtained 
       with the Australia Telescope which is funded by the Commonwealth of 
       Australia for operations as a National Facility managed by CSIRO.}}

\author[B.~Koribalski \& E.~Manthey]
       {B\"arbel Koribalski$^1$ and Eva Manthey$^2$ \\
        $^1$Australia Telescope National Facility, CSIRO, 
            P.O. Box 76, Epping, NSW 1710, Australia\\
        $^2$University of Bochum, Department of Astronomy, 
            44780 Bochum, Germany}

\date{Received date; accepted date}

\begin{document}

\maketitle

\begin{abstract}
We present \HI\ line and 20-cm radio continuum observations of the NGC~1511 
galaxy group obtained with the Australia Telescope Compact Array. The data 
reveal an extended, rather disturbed \HI\ distribution for the peculiar 
starburst galaxy NGC~1511 and a narrow bridge to its small companion galaxy, 
NGC~1511B, which has been severely distorted by the interaction/collision
between the two galaxies. No stellar counterpart to the gaseous bridge has been 
detected. In addition, we find that the peculiar optical ridge to the east of 
NGC~1511 is probably the stellar remnant of a galaxy completed disrupted by 
interactions with NGC~1511. The slightly more distant neighbour, NGC~1511A, 
shows a regular \HI\ velocity field and no obvious signs of interactions.

Radio continuum emission from NGC~1511 reveals three prominent sources on top 
of a more diffuse, extended distribution. We derive an overall star formation 
rate of 7\Msun\,yr$^{-1}$. The most enhanced star formation is found in the 
south-eastern part of the disk, coincident with several bright \HII\ regions,
and closest to the peculiar optical ridge. No continuum emission was detected 
in the companions, but NGC~1511B appears to show an \HII\ region at its faint 
western edge, closest to NGC~1511. The group displays a prime example of 
interaction-induced star formation activity. 

\end{abstract}

 
\section{Introduction} 
Neutral hydrogen gas (\HI) is the best tracer for galaxy-galaxy interactions. 
Since the \HI\ envelope of a galaxy is generally more extended than its stellar
disk (Broeils \& van Woerden 1994; Salpeter \& Hoffman 1996) it is more easily
disrupted by tidal forces from neighboring galaxies. Galaxy pairs and groups 
often show very complex extended \HI\ distributions, incl. \HI\ tails and 
bridges, even if the corresponding optical images show little disturbance. For 
examples see Irwin (1994), Smith, Struck \& Pogge (1997), Yun et al. (1994), 
Koribalski \& Dickey (2004), and the \HI\ Rogues Gallery (Hibbard et al. 2001).
Here we use \HI\ line observations to investigate the interaction history and 
dynamics of the NGC~1511 galaxy group.

The starburst galaxy NGC~1511 is the most prominent member of a compact galaxy 
triplet at a distance of $\sim$17.5 Mpc (\Ho\ = 75\kms\,Mpc$^{-1}$). Its 
companions, NGC~1511B and NGC~1511A, lie at projected distances of 7\farcm3 
(37 kpc) and 10\farcm7 (54 kpc), respectively. Table~1 lists some properties 
for all three galaxies. The stellar disk of NGC~1511 (see, e.g., Eskridge et 
al. 2000, Sandage \& Bedke 1994) contains several bright hotspots, a prominent
dust lane and various peculiar extensions. Faint loops and plumes are seen to 
the northeast of the disk while the hotspots are located in a peculiar hook at 
the south-eastern end of the disk which is best seen in infrared images 
(Eskridge et al. 2000, Jarret et al. 2003). In addition, deep optical images 
reveal an extended vertical ridge $\sim$1\farcm5 east of the south-eastern end 
of the disk, the nature of which we will explore here. Optical short exposures 
of NGC~1511 show a remarkable resemblance to the peculiar starburst galaxy 
NGC~1808 (Koribalski et al. 1993). The optical diameter of NGC~1511 ($3\farcm5 
\times 1\farcm2$ or 17.5 kpc $\times$ 6.1 kpc) is about twice that of the 
neighboring galaxies NGC~1511A and B. While NGC~1511A appears slightly inclined
and symmetric, NGC~1511B is an extremely thin, edge-on galaxy with a faint 
extension to the west. There is a hint of an \HII\ region at the western most 
end. The pecularities of both NGC~1511 and NGC~1511B suggest they are 
interacting. Koribalski (1996) briefly commented on the post interaction/merger
nature of NGC~1511. The \HI\ distribution of the triplet is also briefly 
discussed by Nordgren et al. (1997).
The star formation history of NGC~1511 can be used to estimate the interaction 
time scale. Sekiguchi \& Wolstencroft (1993) find that NGC~1511 has an \HII\ 
region like emission spectrum with high excitation (see also Kewley et al. 
2000), suggesting on-going star formation mixed with a burst $\sim10^8 - 10^9$ 
yrs ago. Thornley et al. (2000) find indications of massive star formation 
($\ga$50\Msun) on a short timescale of $\le 10^{7}$ yr using mid-infrared 
spectroscopy. Elfhag et al. (1996) observed the center of NGC~1511 with the 
SEST and find a CO intensity of $I_{\rm CO}$ = 11.5 K\kms\ (adjusted to our
adopted distance). This corresponds to a nuclear H$_2$ mass of $5.5 \times 
10^8$\Msun\ if we adopt the standard conversion factor of $N_{\rm H_2} / 
I_{\rm CO} = 2.3 \times 10^{20}$ H$_2$ cm$^{-2}$ (K\kms)$^{-1}$ (Strong et 
al. 1988).
Radio continuum, X-ray, and UV data were obtained by Dahlem et al. (2001,
2003), while Lehnert \& Heckman (1995) presented an \Ha\ image.
These data confirm the starburst character of NGC~1511.
 
\begin{table*} 
\caption{Some basic parameters of the observed spiral galaxies.}
\label{tab:table1} 
\begin{tabular}{lcccc}
\hline
                  & NGC~1511 & NGC~1511A & NGC~1511B & Ref. \\
\hline
center position:  \\
~~$\alpha$(J2000) & $03^{\rm h}59^{\rm m}35\fs8$
                  & $04^{\rm h}00^{\rm m}19\fs4$
                  & $04^{\rm h}00^{\rm m}54\fs6$     & (1) \\
~~$\delta$(J2000) & --67\degr38\arcmin07\arcsec
                  & --67\degr48\arcmin28\arcsec
                  & --67\degr36\arcmin42\arcsec            \\
type              & SAa pec  & SB0      &  SBd?      & (1) \\
optical diameter  & $3\farcm5 \times 1\farcm2$
                  & $1\farcm7 \times 0\farcm4$
                  & $1\farcm7 \times 0\farcm3$       & (1) \\
inclination       & $74 \pm ~2$ & $75 \pm ~4$ & $87 \pm ~4$ & (6)\\ 
position angle    & 125\degr  & 110\degr  & 98\degr   & (1,2) \\
$v_{\rm HI}$ [\kms] & $1351 \pm ~7$ & $1323 \pm 10$ & $1304 \pm 3$  & (1,3,6)\\
$v_{\rm opt}$ [\kms]& $1334 \pm 48$ & $1358 \pm 69$ & ---           & (1)\\
$v_{\rm H\alpha}$ [\kms]
                    & $1334 \pm 10$ & $1291 \pm 10$ & $1439 \pm 10$ & (2)\\
$v_{\rm CO}$ [\kms] & $1270 \pm ~7$ &  ---          & ---           & (4)\\
\FHI\  [Jy\kms]           & 73.6  &   20.6  & ---  & (3)\\
S$_{60\mu {\rm m}}$  [Jy] & 23.67 &    0.44 & ---  & (5)\\
S$_{100\mu {\rm m}}$ [Jy] & 37.99 & $<$2.16 & ---  & (5)\\
B magnitude               & 12.11 & 14.23 & 15.12  & (1)\\
${\rm A_B}$               & 0.265 & 0.257 & 0.236  & (7)\\
\LB\ [10$^9$\Lsun]        &  4.9  &  0.7  & 0.3    & \\
\LX\ [10$^{40}$ erg\,s$^{-1}$] & 1.1 & ---  & ---  & (8)\\ 
\hline
\end{tabular}
\flushleft
References: (1) de Vaucouleurs et al. 1991 [RC3], 
            (2) Lauberts 1982 [ESO-LV], 
            (3) Mathewson \& Ford 1996,
            (4) Elfhag et al. 1996,
            (5) Moshir et al. 1990,
            (6) Nordgren et al. 1997,
            (7) Schlegel et al. 1998,
            (8) Dahlem et al. 2003.
\end{table*}

The paper is organized as follows: in Section~2 we summarize the observations
and data reduction. In Section~3 we present the 20-cm radio continuum results 
and derive star formation rates for NGC~1511. The \HI\ morphology and 
kinematics of all members of the NGC~1511 galaxy group are described in 
Section~4. The overall gas dynamics are discussed in Section~5, and Section~6
contains our conclusions.

\section{Observations and Data Reduction} 
\HI\ line and 20-cm radio continuum observations of the NGC~1511 galaxy group 
were obtained with the Australia Telescope Compact Array (ATCA) in March and 
August 1995 using the 1.5A and 375 arrays. For a summary of the observations 
see Table~2.

Data reduction was carried out with the {\sc miriad} software 
package using standard procedures. The data were split into a narrow band 
radio continuum and an \HI\ line data set using a first order fit to the 
line-free channels. The \HI\ channel maps were made using `natural' (na) and
`robust' (r=0) weighting of the {\em uv}-data in the velocity range from 
1050 to 1550\kms\ using steps of 10\kms. The resulting synthesized beams are
$64\arcsec \times 63\arcsec$ (na) and $31.8\arcsec \times 31.2\arcsec$ (r=0); 
the measured r.m.s. noise is 1.7 and 1.8 mJy\,beam$^{-1}$, respectively. The 
\HI\ data cubes were corrected for primary beam attenuation. 
One Jy\,beam$^{-1}$\kms\ corresponds to an \HI\ column density of $2.8 \times 
10^{20}$ atoms~cm$^{-2}$ (na) and $11.2 \times 10^{20}$ atoms~cm$^{-2}$ (r=0).

Radio continuum maps were made using `uniform' weighting of the {\em uv}-data. 
After CLEANing they were restored with a synthesized beam of 6\arcsec, 
resulting in an r.m.s. of 0.5 mJy\,beam$^{-1}$. A low-resolution image was 
created using `natural weighting'; only the shortest 10 baselines were used. 
The beam size here is $76\arcsec \times 71\arcsec$, with an r.m.s. of 0.8 
mJy\,beam$^{-1}$.

\begin{table} 
\caption{Observing parameters.}
\label{tab:table2} 
\begin{tabular}{lcc}
\hline
ATCA configuration    & 1.5A        & 375         \\
\hline
primary beam          & \multicolumn{2}{c}{33\arcmin}  \\
pointing position     & \multicolumn{2}{c}
                        {$03^{\rm h}\,59^{\rm m}\,25^{\rm s}$} \\
~~in $\alpha,\delta$(J2000)& \multicolumn{2}{c}
                        {--67\degr\,47\arcmin\,00\arcsec}     \\
total observing time  & 712 min.    & 647 min. \\
center frequency      & \multicolumn{2}{c}{1414 MHz} \\
total bandwidth       & \multicolumn{2}{c}{   8 MHz} \\
number of channels    & \multicolumn{2}{c}{ 512    } \\
velocity resolution   & \multicolumn{2}{c}{6.6~\kms} \\
\multicolumn{3}{l}{~~~~(after Hanning smoothing)} \\
\multicolumn{3}{l}{Calibrator flux densities:} \\
primary               & \multicolumn{2}{c}{1934--638 (14.88 Jy)} \\
secondary             & \multicolumn{2}{c}{0355--669 (0.97 Jy)} \\
\hline
\end{tabular}
\end{table}

\section{Radio Continuum Results} 
\begin{figure}
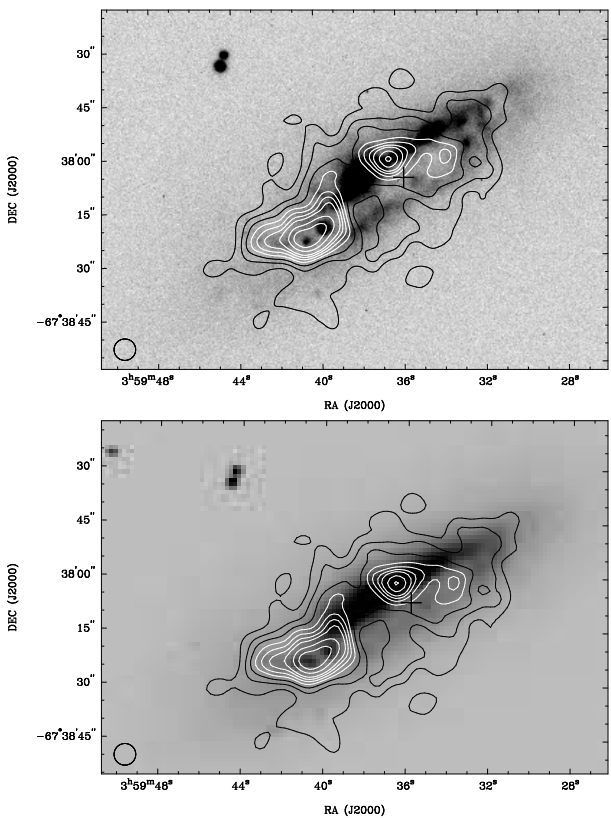
 
\begin{tabular}{c}
  \mbox{\psfig{file=ngc1511.contB2.ps,width=8cm,angle=-90}} \\
  \mbox{\psfig{file=ngc1511.contH2.ps,width=8cm,angle=-90}}
\end{tabular}
\caption{High-resolution 20-cm radio continuum emission of NGC~1511 overlaid 
   onto B-band (top) and H-band (bottom) images from Eskridge et al. (2000). 
   The contour levels are 1, 1.5, 2, 2.5, 3, 3.5, 4, 4.5, 5, 5.5, 6, and 6.5 
   mJy\,beam$^{-1}$. The galaxy center (see Table~1) is marked with a cross.
   The synthesized beam (6\arcsec) is displayed at the bottom left corner.}
\label{fig:cont}
\end{figure}

Since the overall, extended radio continuum emission of NGC~1511 was analysed 
by Dahlem et al. (2001), we concentrate here on the central region. 
Fig.~\ref{fig:cont} shows a high-resolution 20-cm radio continuum 
image of NGC~1511 overlaid onto B- and H-band images from Eskridge et al. 
(2000). We identify three maxima in at $\alpha,\delta$(J2000) = 
$03^{\rm h}\,59^{\rm m}$40\fs7, --67\degr\,38\arcmin\,23\farcs8 (SE complex),
$03^{\rm h}\,59^{\rm m}$36\fs5, --67\degr\,38\arcmin\,02\farcs0 (middle  
source) and
$03^{\rm h}\,59^{\rm m}$33\fs7, --67\degr\,38\arcmin\,01\farcs1 (faint western 
source) with approximate peak flux densities of 53, 27 and 13 mJy\,beam$^{-1}$, 
respectively. The SE complex corresponds to the peculiar hook which is best 
seen in the H-band image (Eskridge et al. 2000) and the 2MASS JHKs-image 
(Jarret et al. 2003). At the same location, the B-band image shows a chain of 
bright hotspots. These are young \HII\ regions (see Lehnert \& Heckman 1995)
consistent with increased star formation in that part of the disk of NGC~1511. 
The middle source, which lies $\sim$6\arcsec\ northeast of the galaxy center 
as given in Table~1, and the faint western source also appear to coincide with 
hotspots in the \Ha\ image.
No radio continuum emission was detected from the companions, NGC~1511A and 
NGC~1511B, with 5$\sigma$ upper limits to their flux densities of $\sim$4 mJy. 
We detect several background sources in the observed field; the two 
brightest are SUMSS\,J035734--674941 (30 mJy) and SUMSS\,J035949--680052 
(68 mJy).

To estimate the star formation rate (SFR) of NGC~1511 from the 20-cm radio 
continuum flux density, $S_{20}$ (Jy), we use SFR ($M_{\odot}$\,yr$^{-1}$) 
= $0.14~D^2~S_{\rm 20\,cm}$ derived from Condon (1992) and Haarsma et al. 
(2000), where $D$ is the distance in Mpc. For NGC~1511 we measure a primary
beam corrected 20-cm radio continuum flux density of 167 mJy (using the 
low-resolution map) and hence a star formation rate of $\sim$7\Msun\,yr$^{-1}$.
There are indications of a faint extended halo, but our data are not sufficient
to study this emission. For comparison, Dahlem et al. (2001) measure a total 
20-cm radio continuum flux density of $156 \pm 8$ mJy. For NGC~1511A and B we 
derive upper limits of SFR $<$ 0.2\Msun\,yr$^{-1}$. 

The star formation rate of a galaxy can also be estimated from its
far-infrared luminosity, $L_{\rm FIR}$, using SFR ($M_{\odot}$\,yr$^{-1}$) 
= 0.17 $L_{\rm FIR}$ (Kennicutt 1998), with $L_{\rm FIR}$ in units of 
$10^9$\Lsun. Using the IRAS 60$\mu$m and 100$\mu$m flux densities given by 
Sanders et al. (1995) (S$_{60\mu{\rm m}}$ = 25.7 Jy, S$_{100\mu{\rm m}}$ = 
41.3 Jy, for comparison see Table~1) we derive a FIR luminosity of $1.36
\times 10^{10}$\Lsun\ for NGC~1511 which results in SFR = 2.2\Msun\,yr$^{-1}$. 
NGC~1511A has an infrared luminosity of $1.4 - 4.0 \times 10^{8}$\Lsun\ and 
SFR = 0.02 -- 0.07\Msun\,yr$^{-1}$. 
No IRAS flux densities are available for NGC~1511B.

Following Helou, Soifer \& Rowan-Robinson (1985) we calculated the parameter
$q$ which is the logarithmic ratio of FIR to radio flux density. For NGC~1511 
we find $q$ = 2.34, consistent with the mean value of 2.3 for normal spiral 
galaxies (Condon 1992).


\section{\HI\ Results} 
\begin{figure*} 
\psfig{file=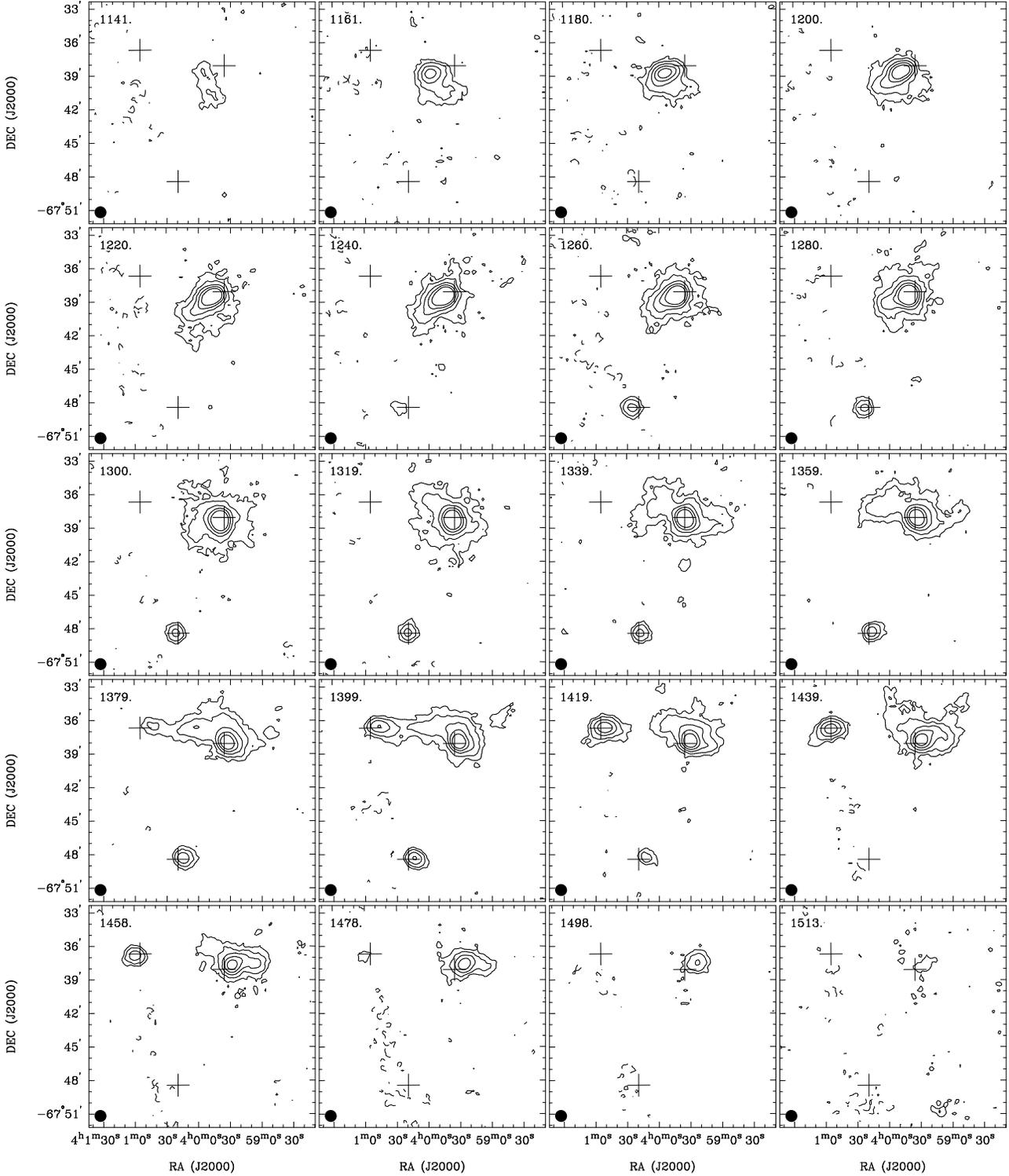,width=17cm,angle=-90}
\caption{\HI\ channel maps of the NGC~1511 group using `natural' weighting. 
  The three galaxy centers are marked as given in Table~\ref{tab:table1}. 
  The contour levels are --3, 3, 6, 12, 20, and 30 mJy\,beam$^{-1}$. The 
  synthesized beam ($64\arcsec \times 63\arcsec$) is displayed at the bottom 
  left corner and the heliocentric velocity of each channel at the top left.}
\label{fig:groupchan}
\end{figure*}

\begin{figure} 
\begin{tabular}{cc}
 \mbox{\psfig{file=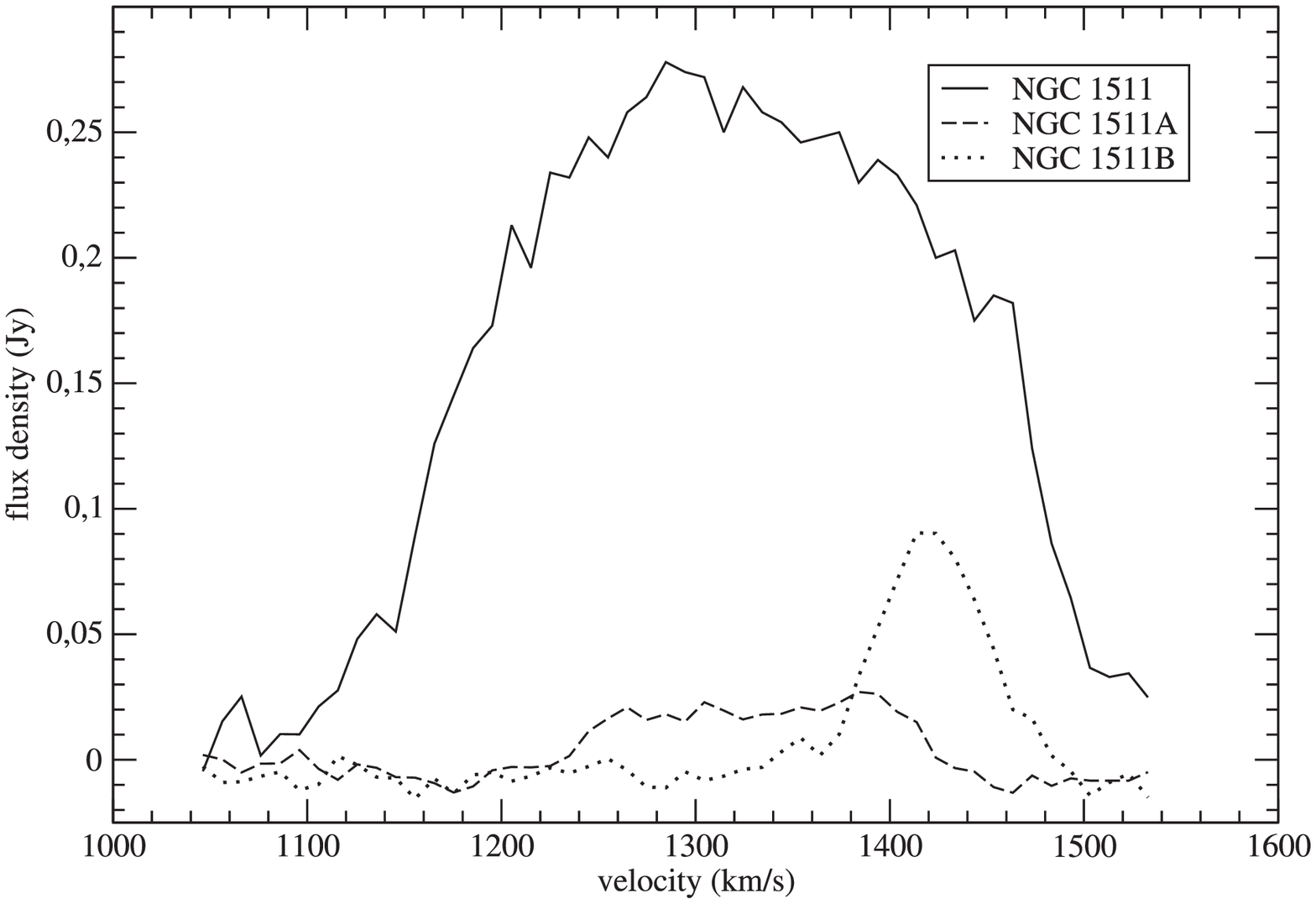,width=8cm}}
\end{tabular}
\caption{Integrated \HI\ spectra of the three galaxies in the NGC~1511 group.}
\label{fig:hispectra}
\end{figure}

The \HI\ channel maps of the NGC~1511 group are shown in 
Fig.~\ref{fig:groupchan}.
All three galaxies are clearly detected, covering a velocity range of 1130 
to 1510\kms. In addition, we detect a thin \HI\ bridge between NGC~1511 and 
NGC~1511B, indicating tidal interactions or a direct collision of the two 
galaxies. No bridge is detected towards NGC~1511A. The individual \HI\ spectra 
are displayed in Fig.~\ref{fig:hispectra}. Most noticeable is the rather narrow
\HI\ spectrum (1370 -- 1480\kms) of the disturbed, edge-on galaxy NGC~1511B, 
in contrast to the broad spectra of NGC~1511 and NGC~1511A. We estimate the 
systemic velocity of NGC~1511B to be 1425\kms, close to the \Ha\ velocity (see 
Table~1). Note that Nordgren et al. (1997) find the \HI\ spectrum of NGC~1511B 
peaks at 1429\kms, in agreement with our value. Because they claim to see 
faint \HI\ emission at lower velocities, they estimate \vsys\ = 1304\kms. The 
\HI\ bridge, which appears to connect the north-western edge of NGC~1511 to 
the western edge of NGC~1511B covers a velocity range from $\sim$1320 to 
1430\kms. Both the \HI\ and the stellar disk of NGC~1511B are affected by the 
interaction. It is likely that the gas in the \HI\ bridge was stripped from 
NGC~1511B by a direct collision with the outer disk of NGC~1511.

\begin{figure*} 
\begin{tabular}{lc}
 \mbox{\psfig{file=ngc1511.mom0.ps,height=7cm,angle=-90}} &
 \mbox{\psfig{file=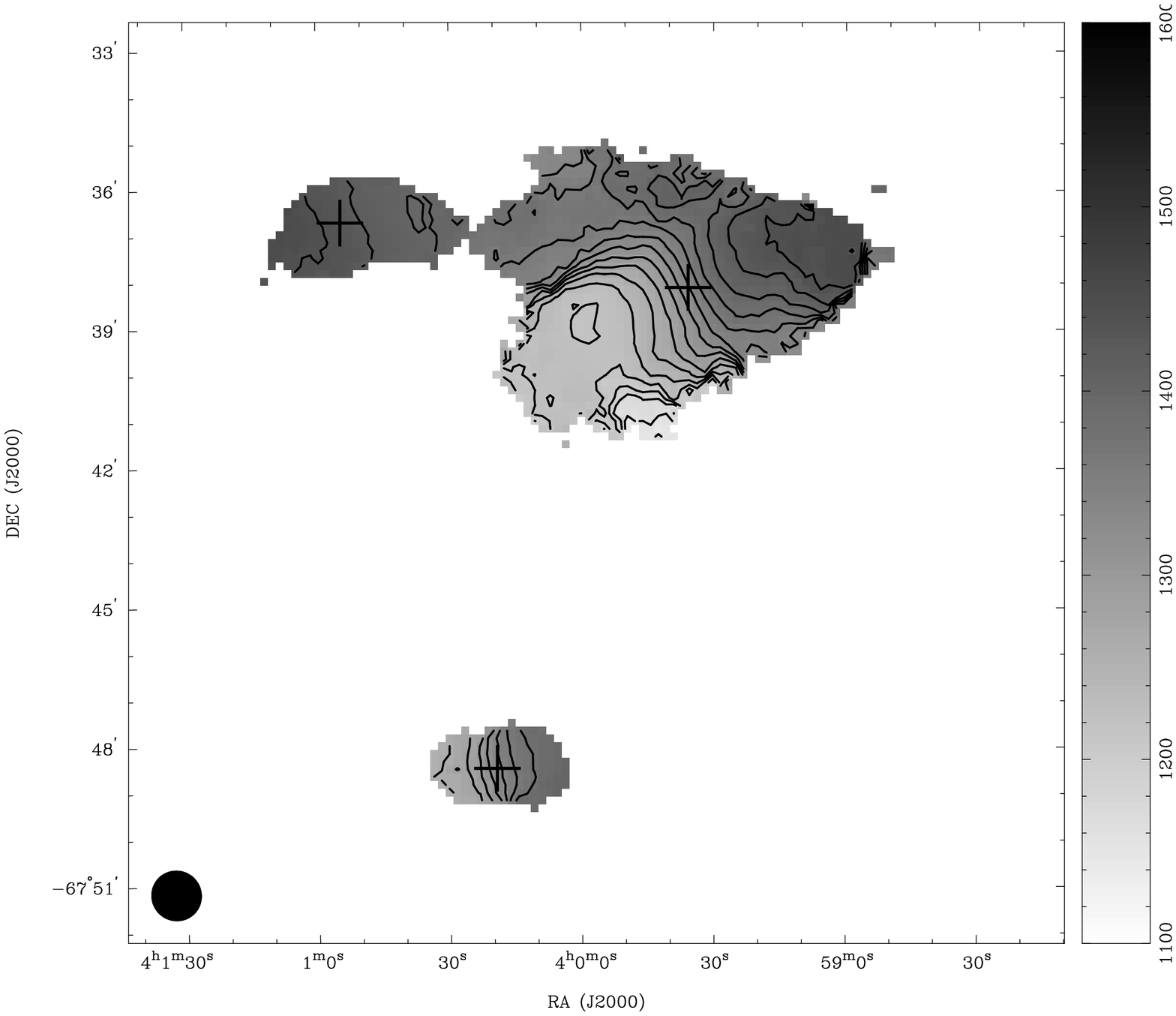,height=7cm,angle=-90}} \\
\mbox{\psfig{file=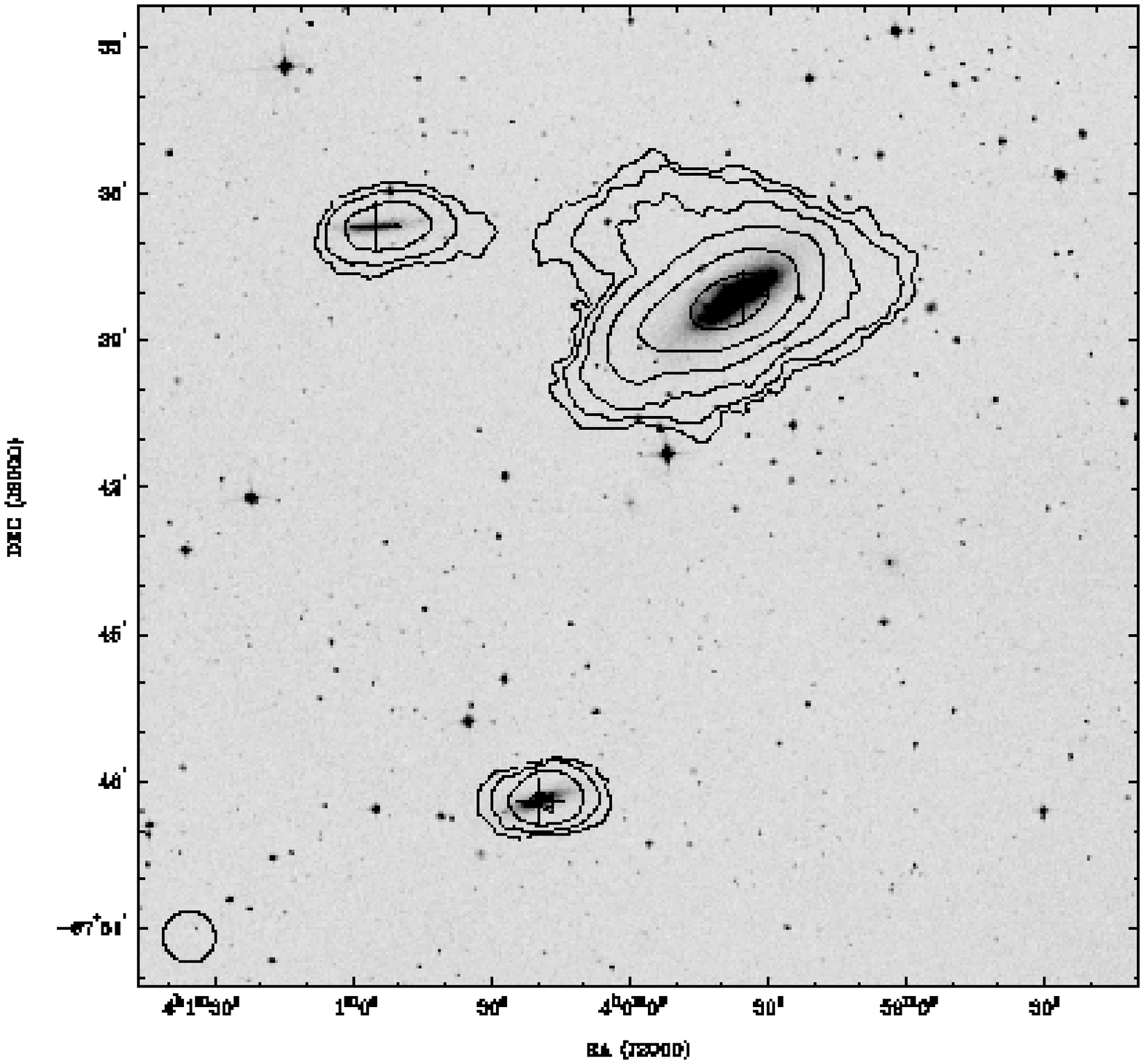,height=7cm,angle=-90}} & 
 \mbox{\psfig{file=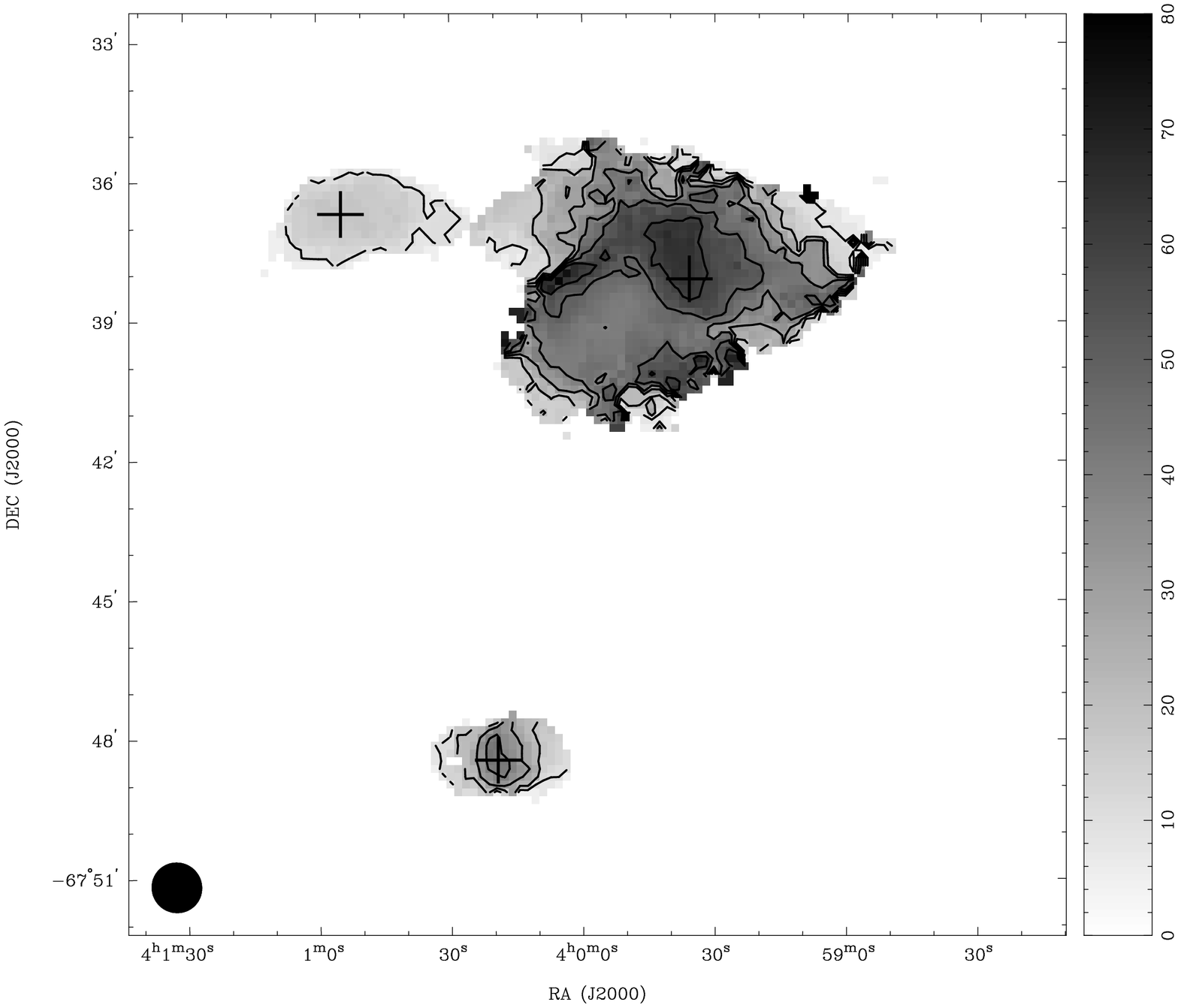,height=7cm,angle=-90}} \\
\end{tabular}
\caption{The \HI\ gas dynamics of the galaxy group NGC~1511 (using `natural'
   weighting). The optical galaxy centers (see Table~1) are marked.
   {\bf --- (top left)} \HI\ distribution (0. moment). The contour levels
   are 0.25, 0.5, 1, 2, 4 and 8 Jy\,beam$^{-1}$\kms\ (corresponding to \HI\ 
   column densities of $6.9 \times 10^{19}$ to $2.2 \times 10^{21}$
   atoms\,cm$^{-2}$). 
   {\bf --- (top right)} Mean \HI\ velocity field (1. moment). The contour
   levels range from 1160 to 1480\kms, in steps of 20\kms.
   {\bf --- (bottom left)} \HI\ distribution (contours) overlaid onto an
   optical image (greyscale) from the Digitised Sky Survey (DSS). The contour 
   levels are as in (a).
   {\bf --- (bottom right)} \HI\ velocity dispersion (2. moment). The contour
   levels are 10, 20, 30, 40, 50, and 60\kms. --- 
  The synthesized beam ($64\arcsec \times 63\arcsec$) is marked in the bottom 
  left corner. }
\label{fig:n1511mom}
\end{figure*}

The \HI\ distribution, the mean \HI\ velocity field and the \HI\ velocity 
dispersion of the NGC~1511 group are shown in Fig.~\ref{fig:n1511mom}.
For a summary of the \HI\ properties see Table~3. 

We measure a total primary-beam corrected \HI\ flux density of \FHI\ = 74 
Jy\kms\ for the NGC~1511 group resulting in an \HI\ mass of \MHI\ = $5.4 
\times 10^{9}$\Msun. For NGC~1511 alone we measure \FHI\ = 66.4 Jy\kms\ 
which corresponds to an \HI\ mass of $4.8 \times 10^9$\Msun, while the two
small companions, NGC~1511A and NGC~1511B, have \HI\ masses of 2.0 and $3.7 
\times 10^{8}$\Msun, respectively. For a summary see Table~3. The HIPASS 
Bright Galaxy Catalog (Koribalski et al. 2004) gives the following \HI\ 
properties for NGC~1511 (HIPASS J0359--67): a systemic velocity of \vsys\ = 
$1333 \pm 5$\kms, 50\% and 20\% velocity widths of 267\kms\ and 323\kms, 
respectively, and \FHI\ = $65.9 \pm 5.0$ Jy\kms, i.e. \MHI\ = 4.8 ($\pm$ 0.3) 
$\times 10^{9}$\Msun. The \HI\ flux density given in HICAT (FHI\ = 74.2\kms,
Meyer et al. 2004) was measured over a larger area ($36\arcmin \times 
36\arcmin$) and is consistent with our \HI\ measurement of the whole group.
This agreement indicates that very little (if any) diffuse \HI\ emission has 
been filtered out by the interferometric observation, i.e. the NGC~1511 group
does not contain large amounts of un-detected \HI\ gas. This is perhaps not
surprising as the main interaction appears to occur between NGC~1511 and 
NGC~1511B and those scales are well covered by the ATCA observations.

Previous \HI\ measurements of NGC~1511, obtained with the 64-m Parkes 
telescope, include \FHI\ = $63.7 \pm 6.1$ Jy\kms\ (Reif et al. 1982),
85.3 Jy\kms\ (Bajaja \& Martin 1985), and 73.6 Jy\kms\ (Mathewson \& 
Ford 1996). 

\begin{table} 
\caption{\HI\ properties of the galaxies and the bridge in the 
    NGC~1511 group as measured with the ATCA.}
\label{tab:table3} 
\begin{tabular}{lccc}
\hline
 Object  &  velocity range &   \FHI   &   \MHI   \\
         &      (\kkms)    & (Jy\kms) & ($10^8$\Msun) \\
\hline
group    & 1130--1510      & 74       &  53.5 \\
NGC~1511 & 1130--1510      & 66.4     &  48.0 \\
NGC~1511A& 1240--1410      &  2.8     &   2.0 \\
NGC~1511B& 1370--1480      &  5.1     &   3.7 \\
\hline          
\end{tabular}
\end{table}

\section{Discussion} 
\subsection{Signatures of Galaxy Interactions}

\begin{figure*} 
\begin{tabular}{c}
\mbox{\psfig{file=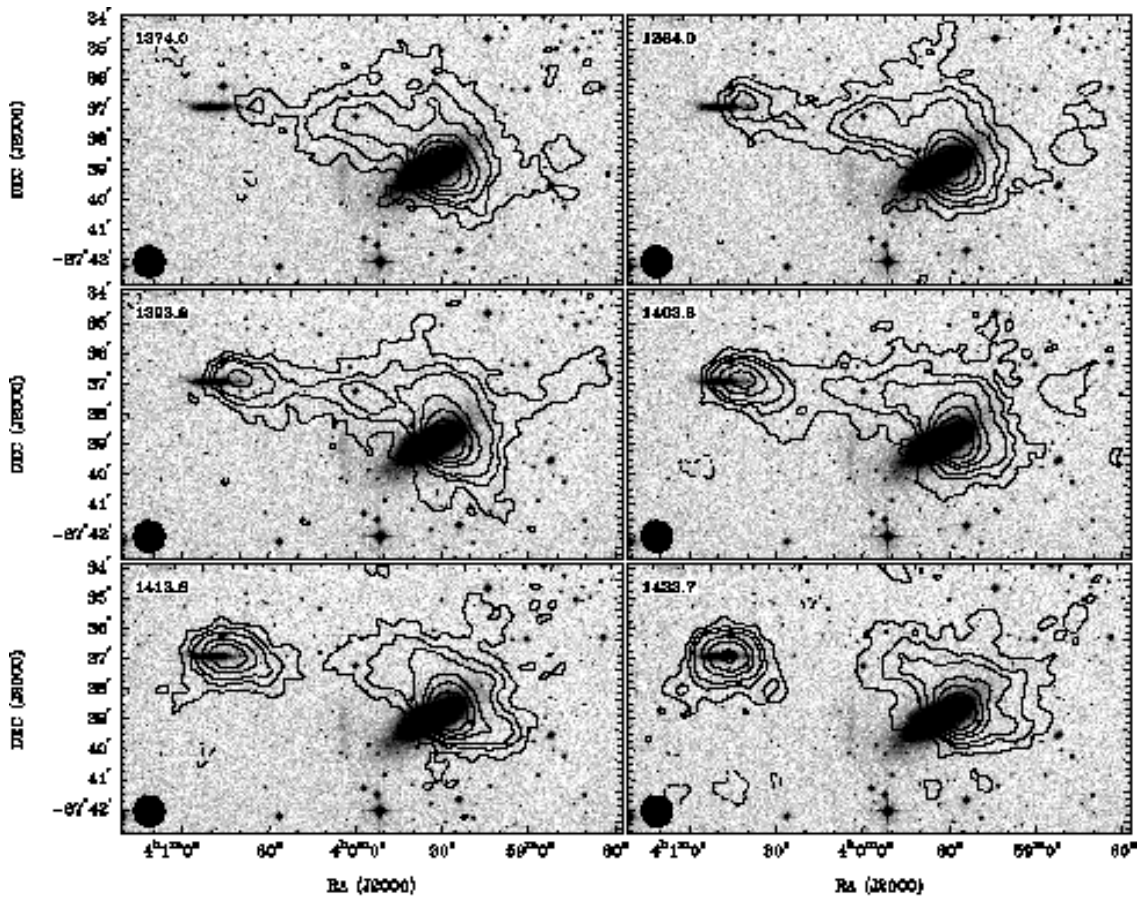,width=15cm,angle=-90}}
\end{tabular}
\caption{\HI\ channel maps (from 1374 to 1424\kms) overlaid onto a high 
   contrast B-band image from the Supercosmos sky survey, emphasizing the 
   narrow bridge between the interacting galaxies NGC~1511 and NGC~1511B. 
   The contour levels are --3, 3, 6, 9, 15, 21, and 30 mJy\,beam$^{-1}$. The 
   beam is displayed at the bottom left corner and the heliocentric velocity 
   of each channel at the top left.}
\label{fig:bridge}
\end{figure*}

Gaseous, intergalactic emission is detected between NGC~1511 and NGC~1511B in 
the velocity range from $\sim$1320 to 1430\kms\ (see Figs.~\ref{fig:groupchan}
and \ref{fig:bridge}). It commences at the north-western side of NGC~1511 and 
appears to curve around to the western side of NGC~1511B, extending over a 
distance of at least 37 kpc. The overall bridge area contains up to $\sim$6
Jy\kms, corresponding to an \HI\ mass of $\sim4 \times 10^8$\Msun; this gas 
is currently attributed to either NGC~1511 or NGC~1511B in Table~3 because,
with the current data, it is impossible to exactly specify how much \HI\ gas 
lies between the galaxies. No stellar counterpart to the bridge has been 
detected in the available optical sky surveys.

\begin{figure*} 
\begin{tabular}{cc}
\mbox{\psfig{file=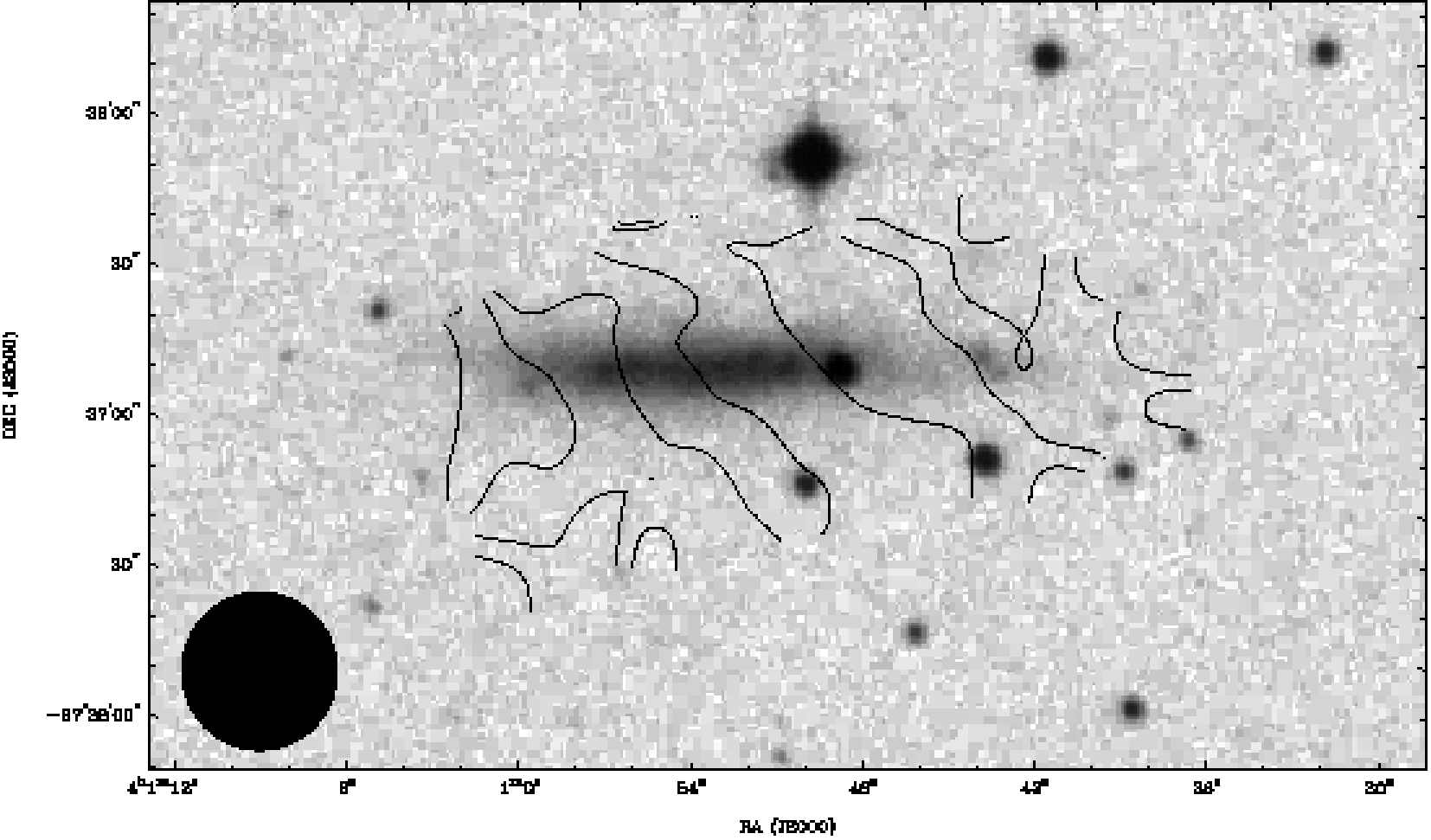,height=4cm,angle=-90}} &
\mbox{\psfig{file=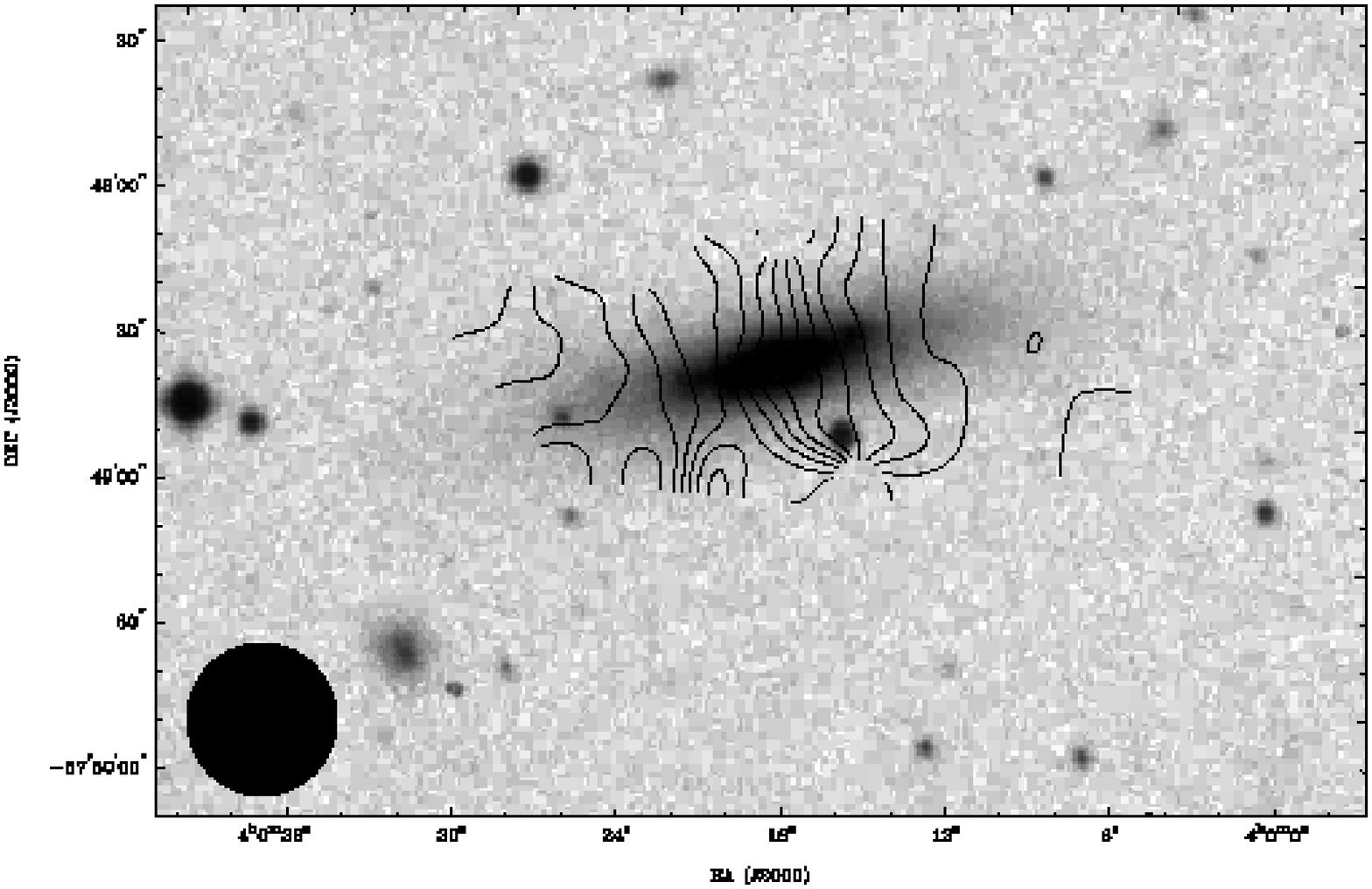,height=4cm,angle=-90}} \\
\end{tabular}
\caption{Mean \HI\ velocity fields of the two companion galaxies, NGC~1511B 
  (left) and NGC~1511A (right) and NGC~1511 (right) overlaid onto B-band 
  images from the Supercosmos sky survey. For these images we used `robust' 
  weighting (r=0), resulting in a synthesized beam of $31.8\arcsec \times 
  31.2\arcsec$, to emphasize the inner disk. The contour levels range from
  1400 to 1460\kms\ (for NGC~1511B) and 1250 to 1400\kms\ (for NGC~1511A),
  in steps of 10\kms. For comparison see Fig.~\ref{fig:n1511mom}.}
\label{fig:velfield2}
\end{figure*}

The stellar distribution of NGC~1511B is extremely thin and edge-on. Its 
main body extends 1\farcm7 (9 kpc) along the major axis. In addition, we 
find a faint extension towards the west with a prominent \HII\ region at
its tip which corresponds to the \HI\ maximum in the 1393\kms\ channel map. 
The \HI\ and the stellar disk of NGC~1511B seem to be elongated and severely 
distorted, especially in the direction of NGC~1511. In addition, the \HI\ 
velocity field of NGC~1511B does not look like that of a normal rotating 
spiral galaxy. The iso-velocity contours are inclined with respect to the 
disk major-axis, at an angle of $\sim$45\degr, and are generally irregular 
(see Fig.~\ref{fig:velfield2}). The velocity dispersion stays around 17\kms\ 
throughout the disk. It seems that the \HI\ gas distribution of NGC~1511B
has been distorted and partially stripped off by gravitational forces during 
the interaction/collision with its massive neighbour, NGC~1511. The dynamical 
mass estimate given for NGC~1511B (see Table~4) is highly uncertain as the 
actual rotational velocity of the galaxy is unknown.

Another most interesting feature found near NGC~1511 is {\em a peculiar
vertical ridge} ($\sim$1\farcm5 in length) only visible in deep optical 
images. It is located $\sim$3\arcmin\ east of the centre of NGC~1511 and
corresponds to a distinct region in the \HI\ velocity field of NGC~1511 (see 
Fig.~\ref{fig:velfield1}). While this region is peculiar with respect to 
the rotation of NGC~1511, it shows remarkably regular iso-velocity contours 
perpendicular to the ridge. We tentatively conclude that what appears like 
a peculiar ridge actually is the stellar remnant of a tidally-disrupted 
galaxy in the NGC~1511 group. \\

Bridges between galaxies, either consisting of gas only or with a stellar 
component, are observed in several galaxy pairs (e.g., Irwin 1994; Smith, 
Struck \& Pogge 1997; Struck \& Smith 2003; Koribalski \& Dickey 2004).
Simulations show two different mechanisms for developing bridges. Toomre \&
Toomre (1972) first predicted bridges to be formed by tidal forces. This
process is successfully modeled by Barnes \& Hernquist (1991) and Mihos \& 
Hernquist (1996), considering an impact parameter larger than the disk radius. 
In this case stars and gas are stripped out in the bridge. Struck (1997)
discusses the formation of ``splash'' bridges which are formed during a 
close encounter (impact parameter smaller than disk radius). Gas clouds of 
both galaxies collide and form a gas bridge between them, typically without 
any stellar counterpart. In this kind of bridge no star formation is initially
expected. The \HI\ bridge between NGC~1511 and NGC~1511B seems to be a 
``splash'' bridge, suggesting that NGC~1511B has plunged through the outer 
disk of NGC~1511 during its closest encounter.

\begin{figure} 
\begin{tabular}{c}
 \mbox{\psfig{file=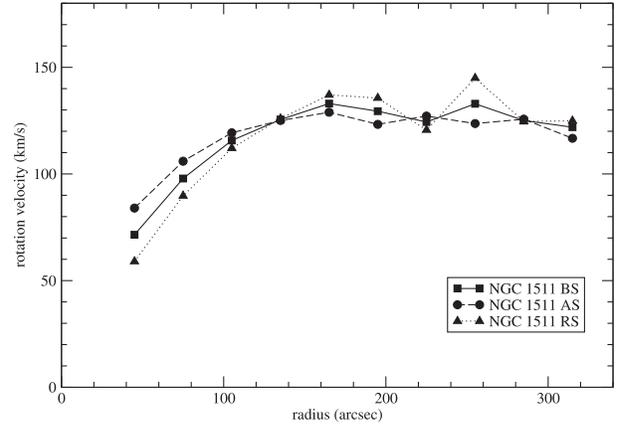,width=8cm}}
\end{tabular}
\caption{Rotation curve model for NGC~1511 for the approaching, receding and
  both sides. For a description of the fit parameters and their uncertainties 
  see Section~5.2.} 
\label{fig:rocur}
\end{figure}

\subsection{The kinematics of NGC~1511 and its companions}

In the inner part of its disk NGC~1511 shows a clear rotation pattern. By 
fitting all rotational parameters simultaneously for radii from 30\arcsec\ 
to 300\arcsec\ we find the kinematic center at $\alpha,\delta$(J2000) = 
$03^{\rm h}\,59^{\rm m}$\,39\fs0 --67\degr\,38\arcmin\,10\farcs0, a systemic 
velocity of \vsys\ = $1333 \pm 5$\kms\ and a position angle of $PA = 300\degr 
\pm 2\degr$. We note that there is a substantial offset between the optical 
center of NGC~1511 as given in Table~\ref{tab:table1}, the 2MASS position, 
and the kinematic center as determined here. Since the fit of the position 
angle appeared constant with radius, we fixed its value to 300$\degr$. The 
inclination angle was difficult to fit, and we used $i = 74\degr$ (Table~1). 
With the two orientation parameters fixed, we derive a rotation curve for the 
approaching side (AS), the receding side (RS), and for both sides (BS) of 
NGC~1511 (see Fig.~\ref{fig:rocur}). While the approaching side of the galaxy 
shows a rather flat rotation curve, the receding side appears slightly 
irregular, possibly showing the effects of interactions which are most evident 
at the higher \HI\ velocities in the system. Given a maximum \HI\ rotation 
velocity of $\sim$130\kms\ and an \HI\ extent of $\sim$300\arcsec\ (25 kpc) 
we derive a dynamical mass of $\sim10^{11}$\Msun. 
Due to the low angular resolution of our \HI\ data compared to the size of
the two companion galaxies, we were not able to fit their rotation curves.

In Table~\ref{tab:table4} we list the systemic velocities, the rotation 
velocities and the \HI\ radii of all three galaxies in the NGC~1511 group,
as well as their dynamical masses and mass-to-light ratios. For NGC~1511A 
and NGC~1511B we estimate \vsys\ as the central velocity of the \HI\ spectrum 
and \vrot\ as the half width of each spectrum, corrected for inclination 
taken from Table~\ref{tab:table1}.
 
\begin{figure*} 
\mbox{\psfig{file=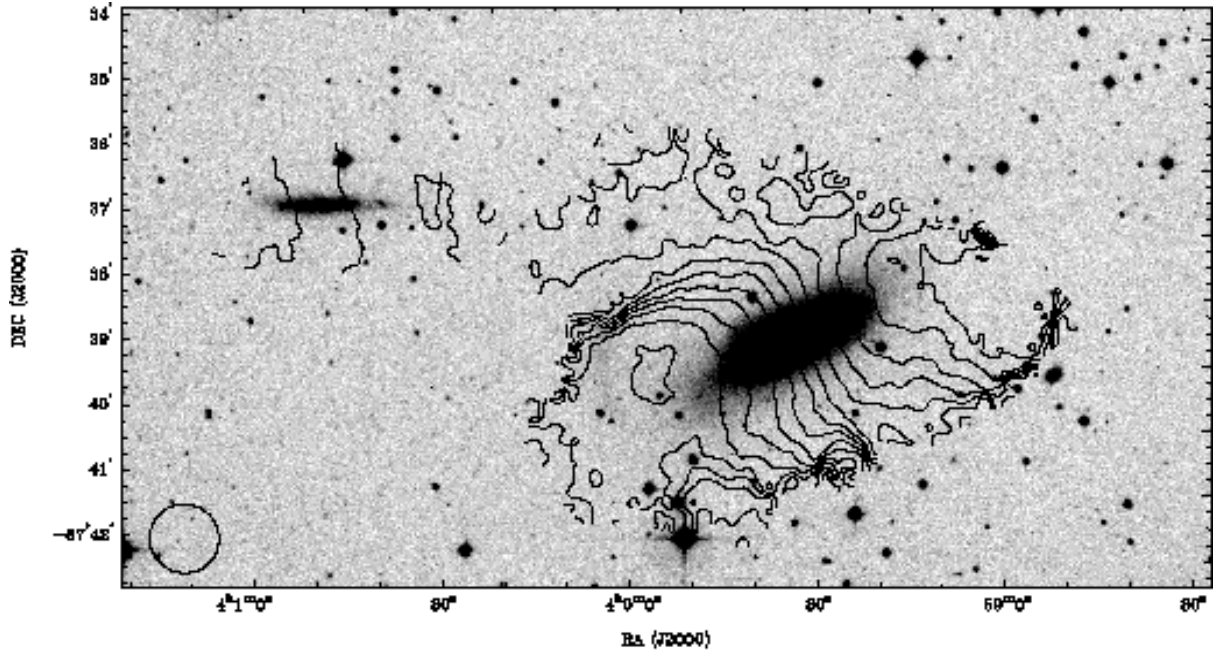,width=16cm,angle=-90}} \\
\caption{Mean \HI\ velocity field of the galaxy NGC~1511 (identical to
  Fig.~\ref{fig:n1511mom}) overlaid onto a B-band image from the Supercosmos 
  sky survey to emphasize the gas kinematics in the region of the peculiar
  optical ridge. The contour levels range from 1160 to 1480\kms, in steps of
  20\kms.}
\label{fig:velfield1}
\end{figure*}

Adding the dynamical masses of the three galaxies together results in 
\Mtot\ (group) = $1.2 \times 10^{11}$\Msun.
For comparison, the virial mass (see Heisler, Tremaine \& Bahcall 1985) of 
the NGC~1511 group is $\sim4.4 \times 10^{11}$\Msun, a factor 3.7 larger.\\
The \HI\ velocity field of NGC~1511 (Fig.~\ref{fig:velfield1}) shows a
broad distribution with an extension to the south-east towards the peculiar
optical ridge. In that area the optical image of NGC~1511 also shows 
peculiarities such as outflows and diffuse extended emission.

\begin{table*} 
\caption{Derived properties of the galaxies in the NGC~1511 triplet.}
\label{tab:table4} 
\begin{tabular}{lccccccc}
\hline
 Object   & \vsys & \vrot & \rHI & \Mtot & \MHI  / \Mtot        
                                         & \Mtot / \LB
                                         & \MHI  / \LB  \\
          &(\kkms)&(\kkms)& (kpc)& ($10^9$\Msun) & & & \\       
\hline
NGC~1511  &  1333 &   130 &   25 & 98 & 0.07 & 20 & 1.4  \\     
NGC~1511A &  1325 &   88  &    8 & 14 & 0.02 & 20 & 0.4  \\     
NGC~1511B &  1425 &   55? &   10 &  7 & 0.07 & 23 & 1.7  \\     
\hline          
\end{tabular}
\end{table*}

\section{Conclusions} 
ATCA \HI\ and 20-cm radio continuum observations of the NGC~1511 triplet
reveal an \HI-rich group with a striking gaseous bridge between the massive 
galaxy NGC~1511 and its minor companion NGC~1511B as well as substantial star 
formation in the disk of NGC~1511. While the gas distribution of NGC1511B is
strongly affected by the interaction, a second, more distant companion galaxy, 
NGC~1511A, remains unaffected. There may have been a fourth member of the 
galaxy group which now appears to be completely disrupted, leaving only the 
faint stellar remnant and a merging \HI\ distribution. In the following we 
briefly summarise our results:

\begin{itemize}
\item \HI\ emission was detected from all three galaxies in the NGC~1511
  triplet. Assuming a distance of 17.5 Mpc for the group we derive \HI\ masses 
  of \MHI\ = $4.8 \times 10^9$\Msun\ (NGC~1511), $2.0 \times 10^8$\Msun\ 
  (NGC~1511A) and $3.7 \times 10^8$\Msun\ (NGC~1511B). 
\item In addition, we found an \HI\ {\em bridge connecting NGC~1511 and 
  NGC~1511B} with an \HI\ mass of up to $4 \times 10^8$\Msun. The bridge, 
  which was most likely formed through a close encounter of the two 
  neighboring galaxies, has no apparent optical counterpart. The projected 
  distance between NGC~1511 and NGC~1511B is $\sim$37 kpc, indicating an 
  approximate time of $\ga$200 Myr since their last collision. 
\item The \HI\ distribution of NGC~1511B appears to have been severely 
  stretched and displaced by the interaction/collision with NGC~1511. It has 
  a rather narrow \HI\ spectrum (for an edge-on galaxy) and an unusual \HI\ 
  velocity field. In contrast, the more distant companion, NGC~1511A, shows 
  no signs of interaction.
\item The {\em peculiar optical ridge} found to the east of NGC~1511 may be 
  the stellar remnant of another group member, now completely disrupted by 
  interactions with NGC~1511.
\item For the NGC~1511 group we derive a total \HI\ mass of $5.4 \times 
  10^{9}$\Msun\ ($\sim$90\% of which resides in NGC~1511 itself), a combined 
  dynamical mass of $\sim1.2 \times 10^{11}$\Msun, and a virial mass of 
  $\sim4.4 \times 10^{11}$\Msun.
\item High resolution ATCA radio continuum data show that most of the emission 
  is occuring in the south-eastern part of the NGC~1511 disk, coincident with 
  a prominent chain of \HII\ regions. We measure a total 20-cm continuum flux 
  density of 167 mJy for NGC~1511 which corresponds to a star formation rate 
  (SFR) of 7\Msun\,yr$^{-1}$. Deriving the SFR from the FIR luminosity we 
  determine a value of 2.2\Msun\,yr$^{-1}$, resulting in a $q$ parameter of 
  2.34. For the companions, NGC~1511A and NGC~1511B, we derive upper limits of 
  SFR = 0.2\Msun\,yr$^{-1}$. 
\end{itemize}

\section*{Acknowledgments}
\begin{itemize}
\item We acknowledge use of the SuperCOSMOS Sky Survey.
\item This research has made extensive use of the NASA/IPAC Extragalactic 
      Database (NED) which is operated by the Jet Propulsion Laboratory, 
      Caltech, under contract with the National Aeronautics and Space 
      Administration. 
\item The Digitized Sky Survey was produced by the Space Telescope Science
      Institute (STScI) and is based on photographic data from the UK Schmidt 
      Telescope, the Royal Observatory Edinburgh, the UK Science and 
      Engineering Research Council, and the Anglo-Australian Observatory.
\item We thank the referee for some excellent comments.
\end{itemize}

\end{document}